\documentclass[journal]{IEEEtran}
\usepackage{cite}
\usepackage[dvips]{graphicx}
\ifCLASSINFOpdf
\else
\fi

\begin{document}

\title{Experimental Determination of \\ the Gain Distribution of \\
an Avalanche Photodiode at Low Gains}

\author{Kenji~Tsujino,
        Makoto~Akiba,
        and~Masahide~Sasaki
% stops a space

\thanks{Manuscript received September 8, 2008; revised October 16, 2008.}% <-this % stops a space
\thanks{K. Tsujino, M. Akiba, and M. Sasaki are with the New Generation Network Research Center, National Institute of Information and Communications Technology, 4-2-1 Nukui-Kitamachi, Koganei, Tokyo 184-8795, Japan. e-mail: (see http://www2.nict.go.jp/w/w113/qit/eng/index.html).}% <-this % stops a space
\thanks{Digital Object Identifier}}

\markboth{Journal of \LaTeX\ Class Files,~Vol.~6, No.~1, January~2007}%
{Shell \MakeLowercase{\textit{et al.}}: Bare Demo of IEEEtran.cls for Journals}

\maketitle

%---------------------------------------
%
%                Abstract
%
%---------------------------------------
\begin{abstract}
\boldmath
A measurement system for determining the gain distributions of 
avalanche photodiodes (APDs) 
in a low gain range is presented. The system is based on 
an ultralow-noise charge--sensitive amplifier 
and detects the output carriers from an APD. 
The noise of the charge--sensitive amplifier is as low as 4.2 electrons 
at a sampling rate of 200 Hz. 
The gain distribution of a commercial Si APD 
with low average gains are presented, 
demonstrating the McIntyre theory in the low gain range.
\end{abstract}


% \fi
\IEEEpeerreviewmaketitle

%---------------------------------------
%
%                Introduction
%
%---------------------------------------
\section{INTRODUCTION}

\IEEEPARstart{G}{ain distribution} is an important characteristic of avalanche photodiodes 
(APDs). The most widely used 
theory for understanding the gain distribution of APDs is that of 
McIntyre\cite{McIntyre1972}. 
In the McIntyre theory, the electron and hole ionization coefficients $\alpha$ and $\beta$ are considered to depend only 
on the local electric field. From this assumption, it is found that the gain 
distribution in the McIntyre theory is considerably different from the Gaussian distribution. The gain distribution for a 
single-carrier injection, for example, has a peak at 
a gain of one for any value of the average gain. 

By using the Monte Carlo simulation to 
investigate the improvement in the avalanche noise performance of 
APDs, Ong et al. found that for a single-carrier injection, 
the peak in the gain distribution of thin p${}^{+}$--i--n${}^{+}$ 
GaAs APDs was shifted 
closer to the average gain\cite{Ong1998}. They showed that at an average gain of 5.1, 
the APD had a peak at a gain of two, whereas conventional APDs have a peak 
at a gain of one. 
These results provide hope for the improvement of the 
photon detection 
efficiency in the detection of single photons when the APDs are operated in the linear 
and sub-Geiger modes\cite{Dautet1993}. If the peak shifts considerably closer 
to the average gain, the APD may have the ability to resolve the photon number\cite{Tsujino2007_1, Tsujino2007_2}.
This behavior is caused by the dead space effect. The dead space is the 
space where impact ionization cannot occur and increases with decreasing 
length of the multiplication region. And because of the dead space effect, 
the randomness of impact ionization process decreases. Then, the gain 
distributions have a peak around the average gain for shorter devices. 
On the other hand, the dead space reduces with excessive electron field 
intensity because the carriers quickly gain sufficient energy to 
initiate an ionization event. That is, this behavior has been found in 
APDs with thin multiplication region ($<$ 1 $\mu$m) and low average gains currently.

Unfortunately, the gain distribution that are expected to exhibit low excess 
noise in contradiction to the McIntyre theory has not been 
experimentally determined so far. Conventional 
charge-sensitive amplifiers that have been widely used to determine 
gain distributions
\cite{Conradi1972,Woodard1994,Dorokhov2003} 
produce noise greater than several hundred electrons, and hence, 
they fail in characterizing the gain distributions of low-noise APDs 
at low average gains. We have also developed 
an ultralow-noise charge-sensitive amplifier  
based on a capacitive transimpedance amplifier (CTIA)\cite{Tsujino2007_1, Tsujino2007_2}; 
however, 
we have never succeeded in determining the gain distribution 
for single-carrier injection
because of the large noise and drift of the 
amplifier\cite{Akiba2005,Tsujino2007_3}. In this letter, 
we demonstrate the determination of the gain distribution for a 
single-carrier injection in low gain ranges 
by improving the noise characteristics of the charge-sensitive amplifier.

%---------------------------------------
%
%          Experimental setup
%
%---------------------------------------
\section{EXPERIMENTAL SETUP}
\begin{figure}[!t]
\centering
\includegraphics[width=3.0in]{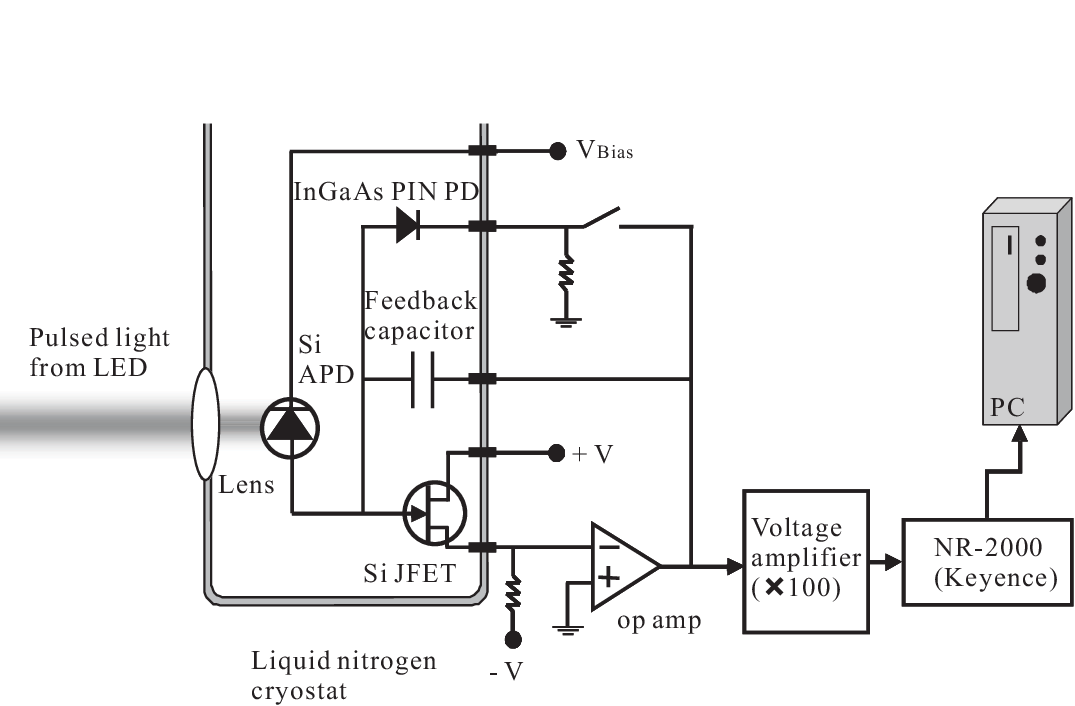}
\caption{Schematic of experimental setup. 
A commercial Si APD was connected to 
an ultralow-noise charge-sensitive amplifier 
that comprised a Si JFET, 
a feedback capacitor, an InGaAs PIN PD for discharging the capacitor
and an operational amplifier (op amp). 
The APD and the circuit elements were cooled at 77 K 
using liquid nitrogen. 
The APD was illuminated with light pulses from a blue light-emitting diode (LED). 
Using a voltage amplifier, 
we amplified the output voltages of the charge-sensitive 
amplifier. 
Finally, we recorded the output voltages in a PC 
using a data collector (NR-2000, Keyence) and then analyzed 
the distribution of the output voltage. }
\label{setup}
\end{figure}

Fig. \ref{setup} shows the experimental setup.
We used a APD (S8890-02, Hamamatsu Photonics) 
with thick multiplication region ($>$ 1 $\mu$m). 
The gain distribution of this type of APD can usually be explained by 
McIntyre noise theory. Then, it would be possible to compare 
the gain distribution measured by our system with the 
predictions of McIntyre noise theory, and hence to test our 
measurement system.     

The CTIA \cite{Tsujino2007_1, Tsujino2007_2} comprised a silicon JFET (Si JFET), 
a feedback capacitor, an InGaAs PIN PD for discharging the capacitor
and an operational amplifier (op amp). 
The Si APD sample was mounted on a SiO$_2$ glass platform 
containing three of the circuit elements of the CTIA (the Si JFET, 
feedback capacitor and InGaAs PIN PD). 
The glass platform was set on the work surface of a cryostat 
and was cooled to 77 K with the aid of liquid nitrogen to suppress 
leak current of semiconductor devices and thermal noise.
The CTIA circuit noise was as low as 4.2 electrons 
at a sampling rate of 200 Hz.

The light signal was produced by a 
450-nm LED, and was focused on the Si APD. 
The wavelength of the LED was chosen to be such that a pure hole 
injection was caused. 
We tried to measure the gain distributions under the pure electron and 
pure hole injection conditions and compared them with the predictions 
of McIntyre noise theory. We found that it is actually the case of hole 
injection at 77 K. 
The signal was modulated to yield pulses with a width of 0.5 ms 
at a repetition rate of 200 Hz. 
To satisfy the conditions for a single-carrier injection, 
the light signal intensity was heavily attenuated 
using neutral density filters 
so that the average number of initial holes 
was $\bar{n} \simeq$ 0.1 per pulse and 
the probability of generation of more than two holes 
was negligible.

The output voltage $V_{out}$ corresponding to $m$ carriers 
generated from the APD is given by
\begin{equation}
V_{out} = \frac{qm}{C_f},
\end{equation}
where $q$ is the elementary charge and 
$C_f$ is the capacitance of the feedback capacitor. 
In our present system, $C_f=$ 0.07 pF 
so that a single output carrier ($m = 1 $) 
induces an output voltage of 2.3 $\mu$V. 
This output voltage is 
further amplified by a manually fabricated voltage amplifier 
with a gain of 100 and a noise level of 0.4 electrons (r.m.s.). 
We digitize the amplified output signals 
using a data acquisition system (NR-2000, Keyence), 
and then store the digitized data in a computer. 
From the stored data, we obtained the gain distribution of the 
APD.

The mean dark current of the APD was 
0.04 electrons/pulse. 
This value was sufficiently less than 
the average number of initial holes. 
We measured the dark current after the warm-up 
drift was stabilized, which took several hours.

%---------------------------------------
%
%   EXPERIMENTAL RESULTS
%
%---------------------------------------
\section{EXPERIMENTAL RESULTS}

\begin{figure}[!t]
\centering
\includegraphics[width=3.0in]{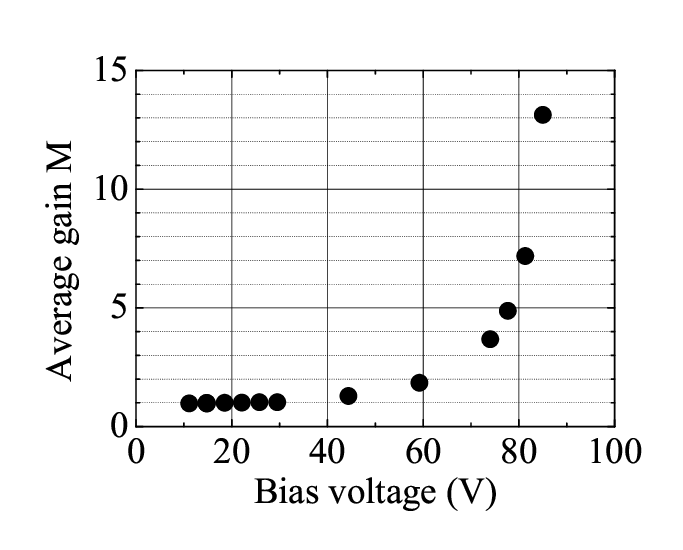}
\caption{Gain vs bias voltage.}
\label{gain}
\end{figure}

Before presenting the gain distribution for a single-carrier injection 
in the low gain range, we discuss the basic characteristics of the 
Si APD sample. Fig. \ref{gain} shows the average gain $M$ curve 
as a function of the bias voltage at 77 K. We measured the gain characteristic 
by using the setup of Fig. \ref{setup} except the average number of initial 
holes was $\bar{n} \simeq$  48 per pulse.
The average gain of the APD is defined to be unity for a bias voltage of 18.5 V 
since the average number of output carriers 
remains constant around this voltage. 
\begin{figure}[!t]
\centering
\includegraphics[width=3.0in]{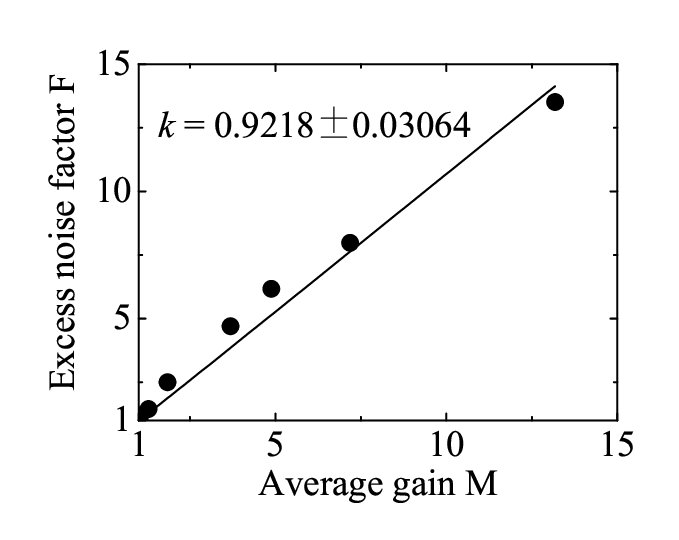}
\caption{Excess noise factor vs. average gain. The solid 
circles are experimental data points. The fitted solid curve 
is a theoretical curve obtained from eq. (3) with $k$ = 
0.9218 $\pm$ 0.03064.}
\label{ENF}
\end{figure}

\begin{figure}[!t]
\centering
\includegraphics[width=3.0in]{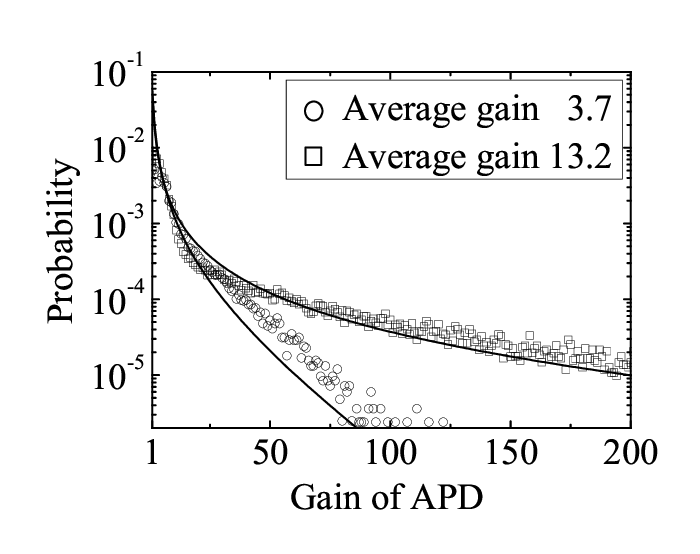}
\caption{Experimental and theoretical gain distributions 
of the S8890 APD at average avalanche gains $M$ of 3.7 and 13.2. 
The average number of initial carriers (holes in this study) is 0.07. 
The noise of the charge-sensitive amplifier is 4.2 electrons. 
The solid curve is the theoretical gain distribution 
obtained from eq. (4)}
\label{GD}
\end{figure}
 The excess noise factor $F$ is
given by

\begin{equation}
F = \frac{\langle G^2 \rangle}{\langle G \rangle^2},
\end{equation}
where, $G$ is a statistical variable that describes the multiplication gain,
and $\langle \cdot \cdot \cdot \rangle$ denote statistical (ensemble) averaging, 
i.g. $\langle G \rangle = M$. We computed the $F$ from analysis of the gain 
distributions obtained by the previous experiment. 
Fig. \ref{ENF} shows the excess noise factor $F$
as a function of the average gain. The obtained data are shown by dots, 
whereas the theoretical fit is shown by 
the solid curve, 
which is obtained from the theory of McIntyre for the hole injection alone 
\cite{McIntyre1966}
\begin{equation}
F(M) = \frac{M}{k} - \left( 2 - \frac{1}{M} \right) 
\left( \frac{1}{k} -1 \right)
\end{equation}
We also estimated the ratio of 
the hole to electron ionization coefficients 
($\beta$ and $\alpha$, respectively) 
as $k = \beta/\alpha=$0.9218 $\pm$ 0.03064.

Fig. \ref{GD} shows the obtained gain distribution at an average gain 
($M$) of 3.7 and 13.2. The average number of initial holes 
was $\bar{n} \simeq$ 0.1 per pulse so that the almost avalanche multiplication 
was caused by single-carrier (here hole) injection. The solid curve is 
the theoretical curve given by 
\begin{equation}
P(x) = p(1) P_{1,m} (k,M),
\end{equation}
where $p(1)$ is the probability of obtaining a single hole in 
a pulse, i.e., $p(1) = \bar{n} \exp(-\bar{n})$. 
$P_{1,m} (k,M)$ is the probability
\begin{eqnarray}
P_{1,m} (k,M) &=& \frac{\left(1-\frac{1}{k}\right)^{m-1} \Gamma \left(\frac{km}{k-1}\right)}
{(m-1)! \left(1-\frac{m-1}{k} \right) \Gamma \left(\frac{k+m-1}{k-1} \right)} \nonumber \\ 
 && \times \left(\frac{M+k-1}{kM}\right)^{\frac{k+m-1}{k-1}} \nonumber \\
 && \times \left( \frac{M-1}{M} \right)^{m-1}
\end{eqnarray}
that an initial single hole will result 
in a total of $m$ holes. 
We can observe fairly good agreement between the measurements and 
the theoretical curve.
This demonstrates that the APDs considered in this study 
can be described well by the McIntyre theory; to the best of our knowledge,
this demonstration is the first in a low gain range.

%
%
%
%
%
%
%---------------------------------------
%
%                Conclusion
%
%---------------------------------------
\section{CONCLUSION}
We have described the measurement system for obtaining the gain 
distributions of APDs in a range of low 
average gains. 
The key technology is an ultralow--noise CTIA 
and detects the carriers generated in an APD. 
We have presented the obtained gain distribution of an Si APD 
for an average gain ($M$) of 3.7 and 13.2 
for a single-hole injection. 

The next important step is to employ our system for the investigation 
of the gain distributions of various types of APDs 
that are expected to exhibit low excess noise in contradiction to
the McIntyre theory\cite{Hu1996, Li1998, Sawada1999, Moszyski2001, Solovov2003, Ikagawa2003}. 
These low-noise APDs are promising devices for use as single-photon 
APDs with high photon detection efficiency since 
the efficiency increases with a decrease in the avalanche multiplication 
noise. Our measurement setup can be a powerful tool for predicting the performance of single-photon APDs.

%---------------------------------------
%
%             Acknowledgment
%
%---------------------------------------
\section*{Acknowledgment}
The authors thank Etsuro Sasaki for technical support.

\ifCLASSOPTIONcaptionsoff
  \newpage
\fi


%---------------------------------------
%
%                references
%
%---------------------------------------


\begin{thebibliography}{1}

\bibitem{McIntyre1972}
R. J. McIntyre, "The distribution of gains in uniformly multiplying avalanche photodiodes: Theory," {\it IEEE Trans. Electron Devices}, vol. ED-19, pp. 703--713, Jun. 1972.

\bibitem{Ong1998}
D. S. Ong, K. F. Li, G. J. Rees, J. P. R. David, P. N. Robson, and G. M. Dunn, "Monte Carlo estimation of 
avalanche noise in thin p${^+}$--i--n${^+}$ GaAs diodes," 
{\it Appl. Phys. Lett.}, vol. 72, pp. 232--234, Jan. 1998.


\bibitem{Dautet1993}
H. Dautet, P. Deschamps, B. Dion, A. MacGregor, D. MacSween, R. McIntyre, C. Trottier, and P. Webb, "Photon counting techniques with silicon avalanche photodiodes," {\it Appl. Opt.}, vol. 32, pp. 3894--3900, Jul. 1993.

\bibitem{Tsujino2007_1}
K. Tsujino, M. Akiba, and M. Sasaki, "Ultralow-noise readout circuit with an avalanche photodiode: Toward a photon-number-resolving detector," {\it Appl. Opt.}, vol. 46, pp. 1009--1014, Mar. 2007.

\bibitem{Tsujino2007_2}
K. Tsujino, M. Akiba, and M. Sasaki, "A charge-integration readout circuit with a linear-mode silicon avalanche photodiode for a photon-number resolving detector," {\it Optics and Spectroscopy}, vol. 103, pp. 86--89, Jul. 2007.

\bibitem{Conradi1972}
J. Conradi, "The distribution of gains in uniformly multiplying avalanche photodiodes: Experimental," {\it IEEE Trans. Electron Devices}, vol. ED-19, pp. 713--718, Jun. 1972.

\bibitem{Woodard1994}
Nathan G. Woodard, Eric G. Hufstedler, and Gregory P. Lafyatis, "Photon counting using a large area avalanche photodiode cooled to 100 K," {\it Appl. Phys. Lett.}, vol. 64, pp. 1177--1179, Mar. 1994.

\bibitem{Dorokhov2003}
A. Dorokhov, A. Glauser, Y. Musienko, C. Regenfus, S. Reucroft, J. Swain, 
"Study of the Hamamatsu avalanche photodiode at liquid nitrogen temperatures," {\it Nucl. Instr. and Meth. A}, vol. 504, pp. 58--61, May. 2003.

\bibitem{Akiba2005}
M. Akiba, M. Fujiwara, and M. Sasaki, "Ultrahigh-sensitivity high-linearity 
photodetection system using a low-gain avalanche photodiode with an 
ultralow-noise readout circuit," {\it Opt. Lett.}, vol. 30, pp. 123--125, Jan. 2005.

\bibitem{Tsujino2007_3}
K. Tsujino, M. Akiba, and M. Sasaki, "Measurement system for acquiring gain distributions of avalanche photodiodes at low gains," {\it Proceedings of SPIE}, vol. 6771, pp. 0Z-1--0Z-8, Sep. 2007.

\bibitem{McIntyre1966}
R. J. McIntyre, "Multiplication noise in uniform avalanche diodes," {\it IEEE Trans. Electron Devices}, vol. ED-13, pp. 164--158, Jan. 1966. 

\bibitem{Hu1996}
C. Hu, K. A. Anselm, B. G. Streetman, and J. C. Campbell, "Noise characteristics of thin multiplication region GaAs avalanche photodiodes," {\it Appl. Phys. Lett.}, vol. 69, pp. 3734--3736, Dec. 1996.

\bibitem{Li1998}
K. F. Li, D. S. Ong, John P. R. David, G. J. Rees, Richard C. Tozer, Peter N. Robson, and R. Grey, 
"Avalanche multiplication noise characteristics in thin GaAs p$^+$--i--n$^+$ diodes," {\it IEEE Trans. Electron Devices}, vol. 45, pp. 2102--2107, Oct. 1998. 

\bibitem{Sawada1999}
K. Sawada, M. Akiyama, and M. Ishida,
"Excess noise characteristics of amorphous silicon staircase 
photodiode films," {\it Appl. Phys. Lett.}, vol. 75, pp. 1470--1472, Sep. 1999.

\bibitem{Moszyski2001}
M. Moszy\'nski, M. Kapusta, M. Balcerzyk, 
M. Szawlowski, D. Wolski, I. W\c egrzecka, and M. W\c egrzecki, 
"Comparative study of avalanche photodiodes with different structures in scintillation detection," {\it IEEE Trans. Nucl. Sci.}, vol. 45, pp. 1205--1210, Aug. 2001.

\bibitem{Solovov2003}
V. N. Solovov, F. Neves, V. Chepel, M. I. Lopes, R. F. Marques and A. J. P. L. Policarpo, 
"Low-temperature performance of a large area avalanche photodiode," {\it Nucl. Instr. and Meth. A}, vol. 504, pp. 53--57, May. 2003.

\bibitem{Ikagawa2003}
T. Ikagawa, J. Kataokaa, Y. Yatsua, N. Kawaia, K. Morib, T. Kamaec, H. Tajimac, T. Mizunoc, Y. Fukazawad, Y. Ishikawae, N. Kawabatae and T. Inutsuka, 
"Performance of large-area avalanche photodiode for low-energy X-rays and $\gamma$ -rays scintillation detection," {\it Nucl. Instr. and Meth. A}, vol. 515, pp. 671--679, Dec. 2003.


\end{thebibliography}
\end{document}